# Evaporative attachment of slow electrons to alkali nanoclusters

Roman Rabinovitch, Chunlei Xia, and Vitaly V. Kresin

*Department of Physics and Astronomy,
University of Southern California, Los Angeles, California 90089-0484*

**ABSTRACT**

The abundance spectrum of $\text{Na}^-_{n\sim7-140}$ anions formed by low energy electron attachment to free nanoclusters is measured to be strongly and nontrivially restructured with respect to the neutral precursor beam. This restructuring is explained in quantitative detail by a general framework of evaporative attachment: an electron is captured by the long-range polarization potential, its energy is transferred into thermal vibrations, and dissipated by evaporative cooling. The data also affirm a formulated relation between the binding energies of cationic, neutral, and anionic clusters, and an adjustment to the prior values of dimer evaporation energies.

Low energy electron attachment to molecules and biomolecules has been extensively explored (see, e.g., the reviews [1-4]) because as an ionization technique it is quite different from electron bombardment or photoionization. Experimental work on electron attachment to clusters has focused on molecular clusters and fullerenes (see, e.g., [1,2,5-9] and references therein).

Our group has studied electron collisions with free sodium nanoclusters in a beam [10-12]. Containing mobile delocalized electrons, alkali clusters are highly polarizable [13-15]. This gives rise to strong long-range forces in interaction with charged particles [16] and to very high electron capture cross sections for $E_{e^-}$ <1 eV. Post-collision anions have been observed [12,17], which raises new basic questions: how is the energy of the captured electron dissipated; what are the relaxation channels; is there an effect on the cluster population? For example, given that the beam of neutral clusters displays "magic numbers" at the electron-shell-closing sizes of $Na_{20,40,58,...}$ [13], will the daughter anions have abundance maxima at these same sizes – which have an enhanced supply of neutral precursors – or will they manage to reorganize in accordance with the shell closing sequence of $Na^-_{19,39,57,...}$? It is worth noting that studies of energy loss by electrons captured by free nanoclusters have parallels with research on the relaxation of carriers injected into size-quantized nanostructures and quantum dots [18].

In principle, a free cluster can dispose of the added electron energy by electron auto detachment, photon radiation, or cluster fragmentation. To gain insight into the actual mechanism, we report on a detailed measurement and analysis of anion mass spectra formed by the attachment process. As will be shown, the entire abundance spectrum undergoes extensive restructuring which can be explained by the statistical mechanism: the energy of the captured electron is promptly thermalized within the cluster, and the latter then cools by evaporating atom and dimer fragments. To our knowledge, this is the first such detailed investigation. It covers a wide cluster size range and renders the description of the entire attachment process – from initial attraction to final rearrangement - complete.

The experiment is outlined in Fig. 1. A beam of neutral clusters from a supersonic expansion source was intersected at a right angle by low energy electrons in the electron gun scattering region. The electrons were produced by a planar dispenser cathode (Spectra-Mat) and constrained into a ribbon-shaped beam by grids, masks, and a collinear 400-gauss magnetic field [10,19,20]. Cluster anions born in the interaction region were extracted with ion optics, focused into a quadrupole mass filter (Extrel QPS9000), and detected by a channel multiplier (DeTech) equipped with a custom-made conversion dynode held at 16 kV. High voltage operation was necessary to efficiently detect the heavy negative ions.

Despite the effort devoted in the setup to the creation and detection of negative ions, the reality is that low-energy sources produce only a limited amount of electron flux. As a result, only ~1% of the original beam became negatively charged [10], and the remainder passed undisturbed into the end chamber, where they were ionized by focused UV light and mass analyzed by a second quadrupole [21]. An essential benefit of this arrangement was that it recorded both the anion (daughter) and the neutral (precursor) cluster mass spectra simultaneously, making it possible to follow and analyze their transformations without distortions due to beam variations.

Because of the aforementioned low ion yield, typical mass-selected anion signals were ~10 counts per second. To raise the beam intensity, we used high source temperatures and carrier gas pressures. The settings of the electron gun were optimized for maximum anion yield. The electron energy distribution was measured by the retarding potential technique [20], giving an effective average electron energy of $E_{e^-}$ =0.1 eV [22]. The corresponding Langevin [23] electron capture cross sections for Na$_{20}$, Na$_{40}$, and Na$_{58}$ are $\sigma_L \approx$1400 Å$^2$, 2100 Å$^2$, and 2500 Å$^2$, respectively.

The anion mass spectra are shown in Fig. 2(a), displaying well-resolved peaks over a wide range. The distribution is significantly restructured relative to that of the neutral clusters, shown for comparison in Fig. 2(b). First, the overall envelope is moved to higher masses. Second, the magic numbers are lowered by one. Third, the relative intensities of the anion peaks between the magic numbers are in an inverse correlation with the intensities of the corresponding neutrals, *which is not a simple pattern shift by one electron number*. While the envelope change can be due to different detection efficiencies of the anions and of the precursor cluster beam, and the magic number shift is intuitively attributable to the extra acquired electron, understanding other nontrivial variations requires a thorough treatment, as outlined below.

The anions are unlikely to be in a metastable surface-bound electron state, because they are detected long (~10$^{-5}$ s) after formation, and are too "floppy" and hot (see below) to sustain such a state. Radiative decay may become important only at collision energies above several eV, i.e., above the collective resonance frequencies [24]. Consequently, all the energy delivered by the captured electron (its initial kinetic energy $E_{e^-}$ plus the cluster electron affinity $EA$) will rapidly dissipate into internal thermal energy of the cluster. If the amount of this energy is sufficient, cluster evaporation will follow. As described below, the "evaporative attachment" [25] picture accounts for the data in detail, without any adjustable parameters. Thus the full electron attachment process is viewed as consisting of three steps.

(I) An electron is captured by the long-range polarization potential of the cluster. The cross section for this process has been found [10,12] to be governed by the aforementioned Langevin formula, $\sigma_L = \left(2\pi^2 e^2 \alpha / E_{e^-}\right)^{1/2}$, where $\alpha$ is the polarizability. This is a relatively smooth function of cluster size [15], hence the initial capture step will not spawn strong size-to-size intensity alterations.

(II) The energy supplied by the electron is rapidly dissipated into heat. The cluster vibrational temperature (the term is employed as a measure of the internal energy contents) increases by $\Delta T = (E_{e^-} + EA)/C$, where the heat capacity of an $n$-atom cluster may be approximated by $C = (3n-6)k_B$.

(III) Clusters cool by evaporating one or more atoms or dimers and losing internal energy with every step [26,27]. The energy loss is $\Delta E = D_n^- + 2k_BT$, where $D_n^-$ is the anion dissociation energy, the second term is the kinetic energy of the outgoing fragment, and $T$ is the cluster temperature prior to evaporation [28]. This proceeds until the temperature drops so much that the evaporation rate becomes negligible on the experimental time scale, and the cluster size distribution at that point then corresponds to the recorded mass spectrum.

The evaporation rate can be expressed as [27,29]

$$r(T) = \tau_0^{-1} n^{2/3} \exp\left(-D_n^- / k_B T^*\right), \qquad (1)$$

where $T^* = T - D_n^-/2C$, the second term reflecting the so-called finite-size correction [27]. Below we take $\tau_0=2.5 \times 10^{-16}$ s, consistent with the value used in Ref. [30] to deduce cluster dissociation energies and with the precise definition discussed, e.g., in Ref. [27]. Eq. (1) makes it clear that the rate is exponentially sensitive to the anion's dissociation energy and vibrational temperature (and thereby to the electron affinity). These quantities do oscillate from one cluster to the next, and the corresponding variations in the evaporative chain are, in actuality, at the root of the observed restructuring of the abundance spectrum. This emphasizes the importance of accounting for thermal effects in interpreting electron capture and transfer reactions for clusters and related systems [31].

The capture of an electron affects the evaporation process in two ways. First, it changes the dissociation energy by altering the number of delocalized electrons in the cluster, i.e., by modifying the electronic shell levels and their occupation. Second, the energy deposited by the electron heats up the cluster and engenders a jump in the evaporation rate. By using Eq. (1) and including both of these factors, the post-attachment evaporation rates for anions can be calculated. The required ingredients are: the internal temperatures of the original neutral clusters, their electron affinities, and the anions' monomer and dimer fragmentation energies and branching ratios.

Applying the analogue of Eq. (1) to the original neutral beam, one can define the temperature $T_n$ for each $Na_n$ cluster based on its lifetime $\tau$ (i.e., flight time from the source) and on $D_n$. This approach is described in detail in Ref. [27], and has been found to be in good agreement with experiment [28,33]. Its result is that $T_n$ may be taken as a flat distribution $F(T)$ spread between $T_n^{min}$ and $T_n^{max}$ with

$$T_n^{max} = D_n/k_B \ln(n^{2/3}\tau/\tau_0) + D_n/2C, \quad T_n^{min} = T_{n+1}^{max} - D_{n+1}/C. \quad (2)$$

The initial temperature of the corresponding anion, $T_n^-$, is found by adding the amount $\Delta T$ specified above, with $EA$ from the $Na_n^-$ photoelectron spectra [34,35]. Typical post-attachment anion average temperatures and spreads range from $520\pm35$ K for $Na_{35}^-$ to $1190\pm180$ K for $Na_7^-$. So the probability for a newly formed anion to evaporate within an elapsed time $t$ is, in view of the flatness of $F(T)$,

$$P_n = \int_{T_{n,min}^-}^{T_{n,max}^-} \left(1 - e^{-r(T)\cdot t}\right) dT, \quad (3)$$

Finally, we need the anion dissociation energies $D_n^-$. Unfortunately, experimentally only cation data, $D_n^+$, are available [30], and for negative ions theoretical values have been computed only for $Na_n^{2-}$ [36]. Using the droplet model [37], the following relation between the dissociation energies of neutral and ionic clusters can be derived, to leading orders in $1/n$ [38]:

$$D_n \approx D_{n+1}^+ - \frac{5}{24n}\frac{e^2}{R_n} - \frac{2}{9n^{4/3}}a_s, \quad D_n^- \approx D_{n+1} + \frac{1}{8n}\frac{e^2}{R_n} - \frac{2}{9n^{4/3}}a_s. \quad (4)$$

Here $R_n = r_0 n^{1/3}$ is the cluster radius, $a_s \approx 1.02$ eV is the surface energy coefficient [37]. (Shell corrections do not enter as long as the number of valence electrons remains unchanged; this is corroborated by jellium model calculations [36].)

Cation dissociations energies were originally determined [30] from rate functions based on unimolecular reaction theory. However, the pre-exponential factor for dimer evaporation did not account for its internal degrees of freedom [39], making it too low by $\approx 10^2$. To correct this, the dimer evaporation energies of Ref. [30] need to be adjusted upward by $\approx 20\%$ [38]. It is satisfying that analysis of our electron capture data, being sensitive to the binding energies, gives support both to this correction and to the relations (4).

Modeling the anion abundance spectra begins with the measured mass spectrum of the neutral precursors, which is convoluted with $\sigma_L$ to generate the parent anion population. Next, evaporative cooling chains are calculated for every cluster. It is important to incorporate both atom and dimer loss pathways [30,40]. The resulting distribution is weighted by an instrumental factor $\sqrt{n}$ (the probability for an ion to exit the electron collision region [20]), and may now be compared with the experimental anion mass spectrum.

This is shown in Fig. 3 for clusters up to $Na_{33}^-$ (the largest size for which dissociation cascades can be fully derived from the dissociation energy data [30]). The calculation matches the experimental pattern well, including such nontrivial features as the aforementioned intensity variations of open-shell clusters, and the peculiar fact that the $Na_{18}^-$ peak is stronger than the closed-shell $Na_{19}^-$ (this is due to extensive evaporation of the parent $Na_{20}^-$).

We conclude that the presented scheme provides an accurate, unified portrait of negative ion formation by low-energy electron attachment: efficient capture by long-range forces, followed by thermalization and by strong rearrangement of the abundance distribution by evaporative emission of atom and dimer fragments. Since evaporative cooling is exponentially sensitive to temperatures and dissociation energies, slow-electron capture offers a useful window into the statistical and binding properties of clusters, as well as molecules with strongly coupled vibrational modes.

Evaporation will be important until nanoclusters grow to be too massive to heat up appreciably. An estimate gives $Na_{n \gtrsim 10^3}$ as the size where the anion mass spectrum will begin to mirror that of the neutral precursor beam. Here, as for stronger-bound clusters and complex molecules heated by electron attachment, thermal radiation [41] will become the dominant release channel. For attachment of more energetic electrons, evaporation will recur.

We thank Dr. K. Hansen and Dr. R. Moro for helpful discussions, D. Boettger for assistance, and J. Ray and DeTech for constructing the high-voltage channeltron. This research was supported by NSF.

**Fig. 1.** Outline of the experiment. A beam of neutral clusters was created by supersonic expansion through a 75 μm nozzle. The source body was kept at 660 °C and the Ar carrier gas pressure varied from 300-600 kPa. The electron gun intersected the cluster beam (collimated to 1.4 mm × 1.4 mm) by a ribbon of slow electrons (1.4 mm × 25.4 mm, ~10 μA). The negatively charged products were extracted with an electrostatic lens, filtered by a quadrupole mass analyzer (QMA), and detected by a channeltron. The remaining neutral clusters were ionized by a UV lamp and detected by another QMA, thus the abundance spectra of precursor and anion clusters were recorded simultaneously.

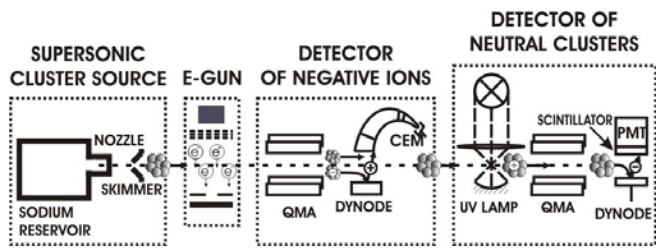

**Fig. 2.** Mass spectra of the electron attachment products (a) and the precursor beam (b). The anion pattern differs from that of the neutral precursors: not only are the magic numbers shifted, but the relative abundances of open-shell clusters are strongly altered. Colors in (a) represent separate segments for which the cluster source was optimized for maximum precursor intensity and the first QMA was adjusted for the strongest mass-resolved anion signal.

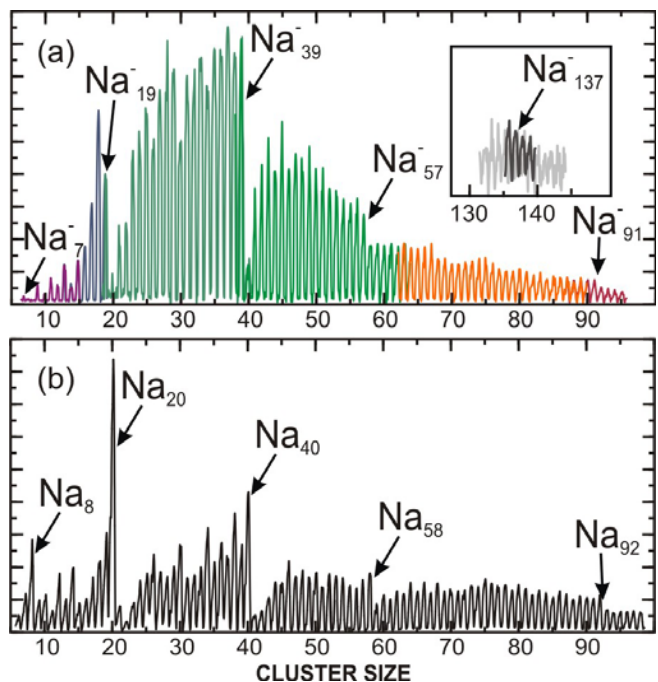

**Fig. 3.** Relative distributions of $Na_n^-$ clusters formed by low-energy electron attachment. The panels correspond to data segments acquired under different optimization conditions. As described in the text, energy released by the captured electron is thermalized, and subsequent cluster cooling via evaporation of atoms and dimers restructures the abundance spectrum. The modeled distribution was derived, without any adjustable parameters, by convoluting the evaporation pathways with the mass spectrum of the precursor neutral beam.

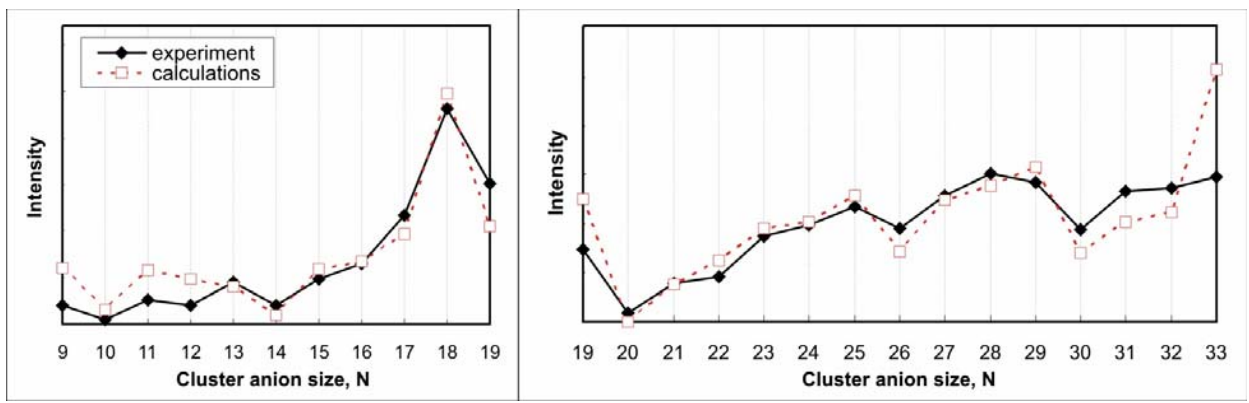